\documentclass[aip,pop,10p,sanserif,reprint,twocolumn]{revtex4-1}
\usepackage[english]{babel}
\usepackage{amsmath, amsthm, amssymb, latexsym, graphicx}
\usepackage[utf8]{inputenc}
\usepackage{color} 
\usepackage{enumerate}
\usepackage{lineno}

\newcommand{\om}{\omega}

\begin{document}

\title{Influence of the Hall effect and electron inertia in collisionless magnetic reconnection}

\author{Nahuel Andr\'es}
\email[E-mail: ]{nandres@iafe.uba.ar.}
\affiliation{Instituto de Astronom\'ia y F\'isica del Espacio, CC. 67, suc. 28, 1428, Buenos Aires, Argentina}

\author{Pablo Dmitruk}
\affiliation{Departamento de Fisica, Facultad de Ciencias Exactas y Naturales, Universidad de Buenos Aires and IFIBA, CONICET, Buenos Aires, 1428, Argentina}

\author{Daniel G\'omez}
\affiliation{Instituto de Astronom\'ia y F\'isica del Espacio, CC. 67, suc. 28, 1428, Buenos Aires, Argentina}
\affiliation{Departamento de F\'isica, Facultad de Ciencias Exactas y Naturales, Universidad de Buenos Aires, Pabell\'on I, 1428, Buenos Aires, Argentina}

\date{\today}

\begin{abstract}
We study the role of the Hall current and electron inertia in collisionless magnetic reconnection within the framework of full two-fluid MHD. At spatial scales smaller than the electron inertial length, a topological change of magnetic field lines exclusively due to electron inertia becomes possible. Assuming stationary conditions, we derive a theoretical scaling for the reconnection rate, which is simply proportional to the Hall parameter. Using a pseudo-spectral code with no dissipative effects, our numerical results confirm this theoretical scaling. In particular, for a sequence of different Hall parameter values, our numerical results show that the width of the current sheet is independent of the Hall parameter while its thickness is of the order of the electron inertial length, thus confirming that the stationary reconnection rate is proportional to the Hall parameter.
\end{abstract}

\maketitle

\section{Introduction}\label{intro}

Magnetic reconnection is a physical process which converts magnetic free energy into kinetic energy and heat. This important mechanism of energy conversion is present in several space environments, such as solar flares and planetary magnetospheres \citep{V1975,D1995,T1996,F2007}. The first model of magnetic reconnection was developed within the framework of one-fluid resistive magnetohydrodynamics (MHD), the so-called Sweet-Parker model \citep{P1957,S1958}. In the Sweet-Parker regime, magnetic resistivity breaks the frozen-in condition at sufficiently small scales, thus allowing magnetic reconnection to occur. In particular, \citet{P1957} showed that the reconnection rate (i.e. the rate of change of magnetic flux due to reconnection) scales as the square root of the magnetic resistivity, which leads to exceedingly low reconnection rates for most space physics environments \citep[e.g.][]{Y2011}. Years later, \citet{P1964} reported a possible way out to the slow-rate problem giving rise to the concept of fast magnetic reconnection, i.e. reconnection rates virtually independent of magnetic resistivity. In contrast to the Sweet-Parker scaling, the Petschek solution only showed a mild (logarithmic) dependence on magnetic resistivity, therefore being considered as fast reconnection. However, numerical results showed that the classical Petschek configuration cannot be attained in simulations with a spatially homogeneus resistivity \citep[e.g.][]{Bi1986}.

For collisionless magnetic reconnection, i.e. where the mean free path of the plasma particles is much larger than any length scale in the system (and therefore the resistivity is negligible), the resistive MHD model is no longer appropriate. From a more general perspective, effects other than magnetic resistivity can break the frozen-in condition, such as electron inertia or non-gyrotropic contributions to pressure. Discussions about the relative importance of electron inertia and non-gyrotropic (off-diagonal) pressure tensor terms can be found elsewhere \citep{Ca1994,He1995,Bi1997,He1998,S1998,Sh2007}. In the present paper we focus on the role of electron inertia. We also assume incompressibility and therefore non-gyrotropic pressure effects are not important \citep{Bi1997}. More specifically, we focus on the physical consequences of including the Hall effect and electron inertia (with isotropic pressures) into a fluidistic description. 

At spatial scales larger than the ion inertial length $\lambda_{i}\equiv c/\om_{pi}$ (where $c$ is the speed of light, $\om_{pi}=\sqrt{4\pi e^2n/m_{i}}$ is the plasma proton frequency, $e$ is the electron charge and $n$ is the plasma density), the MHD description is adequate to describe global phenomena in most of the astrophysical plasmas. However, at scales below $\lambda_i$, where the ions become unmagnetized, the Hall-MHD (HMHD) description becomes valid. At spatial scales of the order of the electron inertial length $\lambda_{e}\equiv c/\om_{pe}$ (where $\om_{pe}=\sqrt{4\pi e^2n/m_{e}}$ is the plasma electron frequency) or smaller, the terms of electron inertia become dominant, and electrons are no longer frozen to the magnetic field lines \citep{V1975}. At this level of description, a topological change of the magnetic field lines exclusively due to electron inertia becomes possible. \citet{A2014a} presented a study of collisionless magnetic reconnection within the framework of Electron Inertia Hall-MHD (EIHMHD), i.e. a two-fluid theoretical framework that extends HMHD and includes the inertia of electrons. Using a pseudo-spectral code with no dissipative effects, the authors numerically confirmed that the change in the topology of the magnetic field lines is exclusively due to the presence of electron inertia. Moreover, they showed that the computed reconnection rates were independent of the mass ratio $m_e/m_i$ and remain a fair fraction of the Alfv\'en velocity, which therefore qualifies as fast reconnection. It is worth mentioning that the level of description of EIHMHD should not be confused with the so called electron MHD (EMHD) approximation. Instead, the EIHMHD model retains the whole dynamics of both the electron and ion flows throughout all the relevant spatial scales. It asymptotically becomes MHD at the largest scales, HMHD at intermediate scales and EMHD at the smallest scales. Under the EMHD approximation, the ions are assumed to be static (because of their much larger mass) and the electrons are the ones to carry the electric current \citep{Bi1997}.

Geospace Environment Modeling (GEM) Reconnection Challenge \citep{Bi2001} was a project designed to study collisionless magnetic reconnection assuming different theoretical approaches such as fully electromagnetic particle in cell \citep{H2001,P2001,S2001}, resistive MHD, HMHD \citep{BH2001,O2001,Ma2001,S2001} and hybrid codes \citep{K2001,S2001}. The authors find that the reconnection rate is insensitive to the mechanism that breaks the frozen-in condition and its particular value is approximately $\sim$0.1 (in dimensionless form). In particular, \citet{S1999} claimed that this values of the reconnection rate is an universal constant as the system become very large.

However, several studies have demonstrated that the reconnection rate might still depend on the value of the Hall parameter \citep{Moa2005,Mob2005,S2008}, the level of turbulent fluctuations \citep{M1986,L1999,Sm2004,Se2009} and the boundary conditions of the problem \citep{W2000,W2001}. This idea that MHD turbulence may play an important role in a magnetic reconnection setup was first proposed by \citet{M1986}. \citet{Sm2004} examined the influence of the Hall effect and level of MHD turbulence on the reconnection rate in 2.5D compressible Hall MHD. Their results indicate that the reconnection rate is enhanced both by increasing the Hall parameter and by the turbulence amplitude. 

Following an approach of single-particle dynamics, \citet{C1985} found an expression for the reconnection rate, which strongly depends on the ion inertia length. \citet{W2000} reported an analytical treatment of quasi-stationary collisionless magnetic reconnection including the Hall effect, scalar electron pressure gradient and electron inertia terms. The authors find that the reconnection rate depends on the ion inertial length, the boundary/initial conditions and the expression for the external driving force. More recently and within the context of incompressible HMHD, \citet{S2008} presented a quantitative analysis of reconnection valid for arbitrary values of the Hall parameter \citep[see also][]{My2008}.

Our main goal in this paper is to study the magnetic reconnection rate, using a full two-fluid model for a completely ionized hydrogen plasma, retaining the Hall current and electron inertia. Within this framework, we calculate a scaling for the quasi-stationary reconnection rate. Our results show that the reconnection rate has a linear dependence on the Hall parameter. In section \ref{eih-mhd} we briefly describe the ideal EIHMHD set of equations. In section \ref{theo} we present our theoretical scaling for the reconnection rate. In section \ref{initcond} we show the set of equations that describes the dynamical evolution of the problem in a 2.5D setup and the corresponding initial conditions. In section \ref{stat_mag_rec}, considering a pseudo-spectral method to accurately run ideal simulations, we present our main numerical results. Finally, in section \ref{conclus} we compare and discuss our results with those reported in the literature and summarize our main conclusions.

\section{Electron Inertia Hall-MHD model}\label{eih-mhd}

The detailed derivation of the EIHMHD model have been presented elsewhere \citep{A2014a,A2014b}. In this section we summarize its key points. The equations of motion for an incompressible plasma made of ions and electrons with mass $m_{i,e}$, charge $\pm e$, density $n_{i}=n_{e}=n$ (because of quasi-neutrality), pressure $p_{i,e}$ and velocity $\textbf{u}_{i,e}$ respectively, can be written as
\begin{eqnarray}\label{motionp}
m_in\frac{d \textbf{u}_i}{dt} &=&  en(\textbf{E}+\frac{1}{c}\textbf{u}_i\times\textbf{B})-\boldsymbol\nabla p_i
\end{eqnarray}
\begin{eqnarray}\label{motione}
m_en\frac{d \textbf{u}_e}{dt} &=&  -en(\textbf{E}+\frac{1}{c}\textbf{u}_e\times\textbf{B})-\boldsymbol\nabla p_e
\end{eqnarray}
\begin{eqnarray}\label{ampere}
\textbf{J}&=&\frac{c}{4\pi}\boldsymbol\nabla\times\textbf{B}={en}(\textbf{u}_i-\textbf{u}_e)
\end{eqnarray}
where 
\begin{equation}
\frac{d\textbf{u}_{e,i}}{dt}\equiv\frac{\partial \textbf{u}_{e,i}}{\partial t}+(\textbf{u}_{e,i}\cdot\boldsymbol\nabla)\textbf{u}_{e,i}
\end{equation}
is the total derivative. Here, $\textbf{B}$ and $\textbf{E}$ are the magnetic and electric fields, $\textbf{J}$ is the electric current density and $c$ is the speed of light. This set of equations can be written in a dimensionless form in terms of a typical length scale $L_0$, the constant particle density $n$, an intensity $B_0$ for the magnetic field, a typical velocity $v_A=B_0/(4\pi nM)^{1/2}$ (the Alfv\'en velocity, where $M\equiv m_i+m_e$) and the electric field in units of $E_0 = v_AB_0/c$,
\begin{eqnarray}\label{dlessp}
 (1-\mu)\frac{d \textbf{u}_i}{dt} &=&  \frac{1}{\lambda}(\textbf{E}+\textbf{u}_i\times\textbf{B})-\boldsymbol\nabla p_i \\ 
\label{dlesse}
 \mu\frac{d \textbf{u}_e}{dt} &=&  -\frac{1}{\lambda}(\textbf{E}+\textbf{u}_e\times\textbf{B})-\boldsymbol\nabla p_e \\ 
\label{dlesso}
\textbf{J} &=& \frac{1}{\lambda} (\textbf{u}_i-\textbf{u}_e) 
\end{eqnarray}
where we have introduced the dimensionless parameters $\mu\equiv m_e/M$ and $\lambda\equiv c/\om_{pM}L_0$ is the dimensionless Hall parameter, and $\om_{pM}=(4\pi e^2n/M)^{1/2}$ has the form of a plasma frequency for a particle of mass $M$. The dimensionless ion and electron inertial lengths can be defined in terms of their corresponding plasma frequencies $\om_{pi,e}=(4\pi e^2n/m_{i,e})^{1/2}$ simply as $\lambda_{i,e}\equiv c/\om_{pi,e}L_0$. Note that in the limit of electron inertia equal to zero, we obtain $\om_{pM}=\om_{pi}$, and therefore $\lambda = \lambda_i = c/\om_{pi}L_0$ reduces to the usual Hall parameter. However, throughout this paper we are going to retain the effect of electron inertia through the parameter $\mu \ne 0$. The expressions for the dimensionless ion and electron inertial scales ($\lambda_{i,e}$) in terms of the two dimensionless parameters $\mu$ and $\lambda$ are simply $\lambda_i=(1-\mu)^{1/2}\lambda$ and $\lambda_e=\mu^{1/2}\lambda$.

For a hydrodynamic description of this two-fluid plasma, we replace the velocity field for each species (i.e. $\textbf{u}_{i,e}$) in terms of two new vector fields. Namely, the hydrodynamic velocity $\textbf{u}$ given by
\begin{equation}\label{hdvelocity}
 \textbf{u} = (1-\mu)\textbf{u}_i+\mu\textbf{u}_e
\end{equation}
and the electric current density $\textbf{J}$ given by \eqref{dlesso}. From equations \eqref{dlesso}-\eqref{hdvelocity}, we can readily obtain the velocity of each species as
\begin{eqnarray}
 \textbf{u}_i&=&\textbf{u} + \mu\lambda\textbf{J} \\
 \textbf{u}_e&=&\textbf{u} - (1-\mu)\lambda\textbf{J}
\end{eqnarray}
The hydrodynamic equation of motion is the sum of the corresponding equations of motion \eqref{dlessp} and \eqref{dlesse} for each species
\begin{equation}\label{hydro}
\frac{d \textbf{u}}{dt} = \textbf{J}\times\left[\textbf{B}-\mu(1-\mu)\lambda^2\nabla^2\textbf{B}\right] - \boldsymbol\nabla p 
\end{equation}
where $p\equiv p_i+p_e$ is the total pressure. Even though most of the terms in equation \eqref{hydro} can easily be identified as a sum of the corresponding terms in equations \eqref{dlessp}-\eqref{dlesse}, the sum of the convective derivatives in these equations are nonlinear terms that give rise to a new nonlinear term in equation \eqref{hydro} which is proportional to $\mu$. Note also that in the limit of negligible electron inertia (i.e., for $\mu\rightarrow 0$), equation \eqref{hydro} reduces to the equation of motion for the traditional one-fluid MHD. This is the case for the Hall-MHD description as well, which is also a two-fluid theoretical description, but considering massless electrons ($\mu=0$). 

On the other hand, the equation of motion for electrons (equation \eqref{dlesse}), using $\textbf{E}=-\partial_t\textbf{A}-\nabla\phi$ and $(\textbf{u}_e\cdot\nabla)\textbf{u}_e=\boldsymbol\omega_e\times\textbf{u}_e+\nabla(u_e^2/2)$ (with $\boldsymbol\omega_e=\nabla\times\textbf{u}_e$ being the electron vorticity) can be written as 
\begin{eqnarray}\label{mote}
\frac{\partial}{\partial t}(\textbf{A}-\mu\lambda\textbf{u}_e) &=& \textbf{u}_e\times(\textbf{B} - \mu\lambda\boldsymbol\omega_e) + \nonumber \\ &+& \nabla(\lambda p_e+\mu\lambda\frac{u_e^2}{2}-\phi)
\end{eqnarray}
We define,
\begin{eqnarray}\label{bp}
\textbf{B}'\equiv \textbf{B}-\mu\lambda\boldsymbol\omega_e&=&\textbf{B}-\mu (1-\mu) \lambda^2\nabla^2\textbf{B} -\mu\lambda\boldsymbol\omega
\end{eqnarray}
where $\boldsymbol\omega=\boldsymbol\nabla\times\textbf{u}$ 
is the hydrodynamic vorticity. Taking the curl of equation (\ref{mote}) we obtain a dynamical equation for the magnetic field 
\begin{eqnarray}\label{dynB}
\partial_t~\textbf{B}' &=& \boldsymbol\nabla\times{[\textbf{u}-(1-\mu)\lambda\textbf{J}]\times\textbf{B}'}
\end{eqnarray} 
Equations \eqref{hydro} and \eqref{dynB} are the EIHMHD equations. It is interesting to note that the presence of the electron mass introduces higher order derivative terms. This certainly has an impact at large wavenumbers, affecting the distribution of energy at very small scales. Note that in the limit of negligible electron inertia (i.e., for $\mu\rightarrow0$), equations \eqref{hydro} and \eqref{dynB} reduce to the standard equation of motion and induction equation of HMHD \citep{G2008,G2013}.

\section{Theoretical scaling of the magnetic reconnection rate}\label{theo}

In the context of collisionless magnetic reconnection, the reconnection region develops a multi-scale structure in which the ion and electron inertial lengths $\lambda_{i,e}$ play a role \citep{Bi1997}. As we discussed in the Introduction, ions can be considered approximately static and electrons are the ones to carry most of the electric current. Also, at these scales the terms of electron inertia become dominant, and the electrons can no longer be frozen-in to the magnetic field lines \citep{V1975}. Therefore, at this level of description, a change in the topology of the magnetic field lines which is exclusively due to electron inertia, becomes possible.

\begin{figure}
\centering
\includegraphics[width=.45\textwidth]{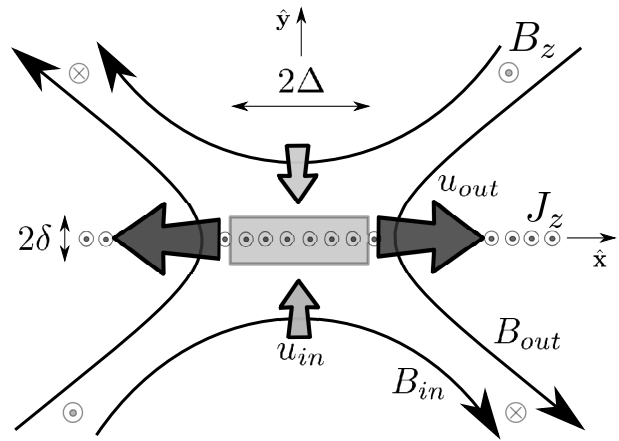}
\caption{Schematic 2.5D reconnection region.}
\label{scheme}
\end{figure}

Within scales near the X-point, where $|\textbf{u}_i|<<|\textbf{J}/en|$, we obtain a scaling for the reconnection rate as a function of $\lambda$ and $\mu$  which are the main parameters of the problem. We consider a rectangular reconnection region with a width $2\delta$ and a length $2\Delta$ (see Figure \ref{scheme}).
By definition, the reconnection rate in a 2D configuration is the out-of-plane component of the electric field (i.e., $E_{z}$) at the X-point. The electric field can be obtained from the ideal equation of motion for the electrons \eqref{motione} as
\begin{eqnarray}\label{electric}
 \textbf{E} =  -\frac{m_e}{e}\bigg[\frac{\partial\textbf{u}_e}{\partial t}+\textbf{u}_e\times(\boldsymbol\omega_e+\frac{e}{m_ec}\textbf{B}) + \nonumber \\ + \boldsymbol\nabla\left(\frac{u_e^2}{2}+\frac{p_e}{m_en}\right)\bigg].
\end{eqnarray}
Under the assumption of quasi-stationarity (i.e., $\partial_t\sim0$) for a 2.5D setup (i.e., $\partial_z\sim0$), the out-of-plane component of the electric field (the $\hat{\textbf{z}}$ direction) reduces to
\begin{eqnarray}
{E}_z = -\frac{m_e}{e}~\hat{\textbf{z}}\cdot\textbf{u}_e\times\boldsymbol\omega_e = \frac{m_e}{e^3n^2}~\hat{\textbf{z}}\cdot\textbf{J}\times(\boldsymbol\nabla\times\textbf{J})
\end{eqnarray}
where we have assumed $\textbf{u}_e\sim-\textbf{J}/en$. 

In view of the sketch shown in Figure \ref{scheme}, close to the X-point is $\partial_x\sim\Delta^{-1}$, $\partial_y\sim\delta^{-1}$ and $J_z=cB_{in}/4\pi\delta$, where $B_{in}$ is the magnetic field at the edge of the reconnection region in the inflow direction. Therefore,
\begin{eqnarray}
{E}_z = \frac{m_e}{e}\left(\frac{c}{4\pi ne}\right)^2\frac{B_zB_{in}}{\Delta\delta^2}.
\end{eqnarray}
To estimate the out-of-plane component of the magnetic field ($B_z$), we consider the $\hat{\textbf{z}}$ component of the curl of equation \eqref{electric} (under quasi-stationary conditions), i.e.
\begin{equation}
\hat{\textbf{z}}\cdot\boldsymbol\nabla\times\left[\textbf{J}\times\left(\frac{e}{cm_e}\textbf{B}-\frac{1}{en}\boldsymbol\nabla\times\textbf{J}\right)\right]=0
\end{equation}
which, in 2.5D setup leads to
\begin{equation}
\textbf{B}_\perp\cdot\boldsymbol\nabla_\perp J_z = \left(\frac{c}{\omega_{pe}}\right)^2\textbf{J}_\perp\cdot\boldsymbol\nabla_\perp(\nabla^2B_z)
\end{equation}
and therefore
\begin{equation}\label{Bz}
B_z=\frac{\omega_{pe}\delta}{c}B_{in}.
\end{equation}
The $\hat{\textbf{z}}$-component of the electric field at the X-point is then
\begin{equation}
E_z=\frac{c}{4\pi en}\frac{B_{in}^2}{\Delta\delta\omega_{pe}}.
\end{equation}
The dimensionless reconnection rate, i.e. $r \equiv c{E}_z/B_0v_A$, becomes
\begin{eqnarray}\label{sca}
r = \frac{c}{\omega_{pM}\Delta}\frac{c}{\omega_{pe}\delta}\left(\frac{B_{in}}{B_0}\right)^{2}
\end{eqnarray}
As it was discussed in the Introduction, we expect $B_{in}$ and $\Delta$ not to depend on $\lambda$ \citep{S2008}. Their particular values are only determined by the boundary and initial conditions. Nevertheless, in the next section we evaluate the potential dependence of $B_{in}$, $\delta$ and $\Delta$ with the Hall parameter in our numerical results.

Assuming that the thickness of the current sheet is essentially the electron inertial length, i.e. $\delta\sim c/\omega_{pe}$ and also that the typical magnetic field intensity is $B_0=B_{in}$ and the typical length scale is $L_0=\Delta$ we obtain
\begin{eqnarray}\label{scaling}
r = \lambda
\end{eqnarray}
Note that if  $\delta\sim c/\omega_{pe}$, according to \eqref{Bz} we also obtain that (in the regime of quasi-stationary reconnection) $B_z\sim B_{in}$. Note also that the reconnection rate is independent of the mass ratio $\mu$, as shown in \citet{A2014a}.

\section{Numerical Results}

\subsection{2.5D Setup and initial conditions}\label{initcond}

In a 2.5D setup, the vector fields depend on two coordinates, say \textit{x} and \textit{y}, although they have their three components. Considering the incompressible case, i.e. $\boldmath\nabla\cdot\textbf{u}=0$, we can write the magnetic and velocity fields as
\begin{eqnarray}\label{Bmag}
\textbf B &=& \boldsymbol\nabla\times[\hat{\textbf z}~a(x,y,t)] + \hat{\textbf z}~b(x,y,t)\\ 
\textbf u &=& \boldsymbol\nabla\times[\hat{\textbf z}~\varphi(x,y,t)] + \hat{\textbf z}~u(x,y,t) 
\end{eqnarray}
where $a(x,y,t)$ and $\varphi(x,y,t)$ are the scalar potential for the magnetic and velocity fields respectively and $b(x,y,t)$ and $u(x,y,t)$ are simply the corresponding out-of-plane components. In terms of these scalar potentials, equations \eqref{hydro} and \eqref{dynB} take the form
\begin{eqnarray}\label{1.1}
\partial_t~\omega &=& [\varphi,\omega] - [a,j] - (1-\mu)\mu\lambda^2[b,\nabla^2b]  \\ \label{1.2}
\partial_t~u &=& [\varphi,u] - [a,b] - (1-\mu)\mu\lambda^2[j,b] \\ \label{1.3}
\partial_t~a' &=& [\varphi - (1-\mu)\lambda b,a']  \\ \label{1.4}
\partial_t~b' &=& [\varphi - (1-\mu)\lambda b,b'] + [u - (1-\mu)\lambda j,a'] 
\end{eqnarray}
where 
\begin{eqnarray}
\omega&=&-\nabla^2\varphi \\
j&=&-\nabla^2a \\
a'&=&a+(1-\mu)\mu\lambda^2j-\mu\lambda u \\
b'&=&b-(1-\mu)\mu\lambda^2\nabla^2b-\mu\lambda\om
\end{eqnarray}
and the nonlinear terms are the standard Poisson brackets, i.e. $[p,q]=\partial_xp\partial_yq-\partial_yp\partial_xq$. The set of equations \eqref{1.1} - \eqref{1.4} describe the dynamical evolution of the magnetic and velocity fields in 2.5D. When $\mu=0$ (massless electrons) this set of equations reduces to the incompressible 2.5D HMHD equations \citep{G2008}.

In the present paper, we performed 2.5D EIHMHD~ simulations using a pseudo-spectral code, which yields exponentially fast numerical convergence and negligible numerical dissipation. The accuracy of the numerical scheme can be verified in part by looking at the behavior of the ideal invariants of the EIHMHD equations in time. The simulations reported here correspond to zero viscosity and resistivity, and the total energy \citep{A2014a,K2014} is conserved by the numerical scheme with an error $\Delta E/E$ of less than $10^{-8}$. The ion and electron helicities were initially zero, and throughout their evolution differ from zero in less than $10^{-15}$. Therefore, hereafter we assume that our code conserves energy. The reconnection processes that are observed to occur, must then be the exclusive result of electron inertia.

Our initial condition to simulate a thin current sheet is given by (assuming periodic boundary conditions in a $2\pi\times2\pi$ box)
\begin{eqnarray}\label{init-cond}
\textbf{B}(x,y,t=0) = \text{B}_0 \bigg[\tanh\left(\frac{y-\frac{3\pi}{2}}{2\pi l}\right) - \nonumber \\ - \tanh\left(\frac{y-\frac{\pi}{2}}{2\pi l}\right) +1\bigg]\hat{\textbf{x}} 
\end{eqnarray}
where, in normalized units, we have $\text{B}_0=1$ and $l=0.02$. To drive reconnection, a monochromatic perturbation $\delta\textbf{B}=\boldsymbol\nabla\times[\hat{\textbf z}~\delta a(x,y)]$ with $\delta a(x,y) = a_0\cos(k_xx)$, $k_x = 1$ and an amplitude of $a_0=0.02\text{B}_0$ is added to the initial condition \eqref{init-cond}. It is worth mentioning that in our simulations we do not use any external driving force, and therefore the reconnection process can be regarded as self-driven. We perform numerical simulations with a spatial resolution of $2048^2$ grid points. For all the runs we use a value of electron to proton mass ratio $m_e/m_i = 0.015$ and different values of the Hall parameter $\lambda$.
\begin{figure}
\centering
\includegraphics[width=.49\textwidth]{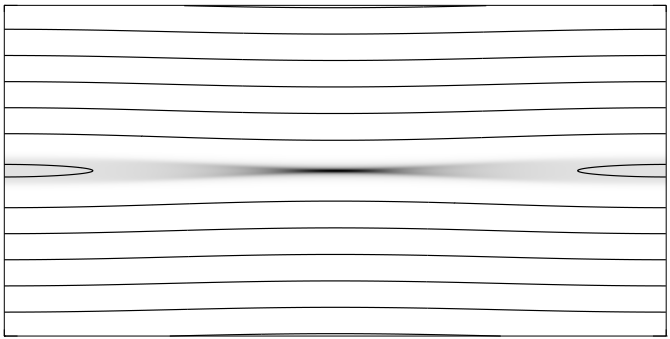} 
\caption{The image (in grayscale) shows the spatial distribution of current density $j(x,y)$ at $t=1.0$ for $\lambda = 0.1$ and $m_e/m_i = 0.015$. Contour levels of $a(x,y)$ are superimposed (black lines).}\label{jj}
\end{figure}
Figure \ref{jj} shows the set up of magnetic reconnection for $\lambda = 0.1$. Contour levels of magnetic flux $a(x,y)$ are in black lines, superimposed to the electric current density component along the $z$ direction, $j(x,y)$, at time $t=0.6$ (in grayscale). We only show half a box of integration for each case, of size $2\pi\times\pi$.

\subsection{Quasi-stationary magnetic reconnection}\label{stat_mag_rec}

Within the framework of EIHMHD, we study the collisionless magnetic reconnection problem varying the dimensionless Hall parameter $\lambda$. Using the initial conditions described in subsection \ref{initcond}, we performed ten ideal runs with a spatial resolution of $2048^2$ grid points for different values of the Hall parameter. Our runs span the range $\lambda=0.07$ to $\lambda=0.16$, with a step of 0.01. The values of $\lambda$ are sufficiently small, to minimize the potential influence of boundary conditions. In all these runs the electron to ion mass ratio corresponds to $m_e/m_i=0.015$. 

\begin{figure}
\centering
\includegraphics[width=.5\textwidth]{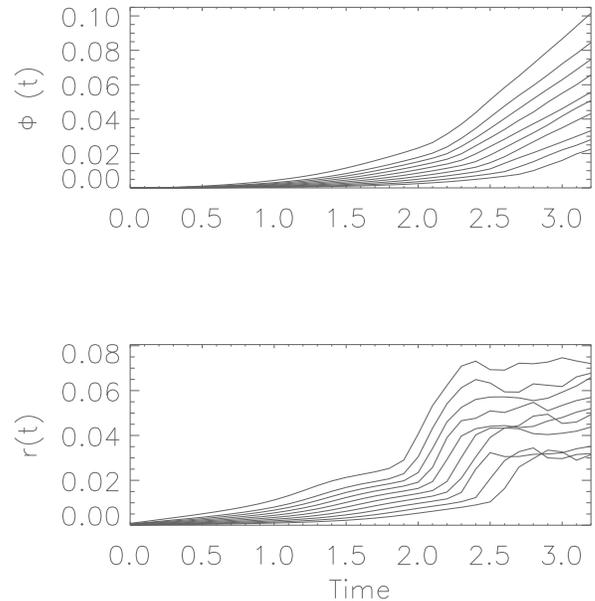}
\caption{Reconnected flux $\Phi$ (upper panel) and reconnection rate $r$ (lower panel) as a function of time for $\lambda=0.07,\cdots,0.16$ (from bottom to top). For all runs the electron to ion mass ratio is $m_e/m_i=0.015$.}
\label{ratefluxall}
\end{figure}

To measure the efficiency of the magnetic reconnection process, the dimensionless reconnection rate $r(t)$ is defined, which is the rate at which magnetic flux flows into the X-point. Using equation \eqref{Bmag} it is straightforward to show that the total reconnected flux $\Phi(t)$ is $\Phi(t) = a_{max} - a_{min}$ \citep{Sm2004,A2014a}. Therefore, the reconnection rate $r(t)$ is the variation of the magnetic flux per unit time, i.e. $r(t)=d\Phi(t)/dt$. Figure \ref{ratefluxall} shows the reconnected flux (upper panel) and reconnection rate (lower panel) as a function of time, for the ten values of the Hall parameter. In contrast to previous claims \citep{S1999,Bi2001}, Figure \ref{ratefluxall} shows that the reconnection rate strongly depends on $\lambda$ and is not a universal constant. As it can be seen, the reconnected flux monotonically increases with $\lambda$, which ultimately leads to an increment of the maximum magnitude of reconnection rate. We also note that as we increase $\lambda$, the maximum reconnection rate occurs at earlier times. Similar behavior has been reported in the literature when the Hall effect is included in Ohm's law \citep{Sm2004,Mob2005,Mo2006}.

\begin{figure}
\centering
\includegraphics[width=.45\textwidth]{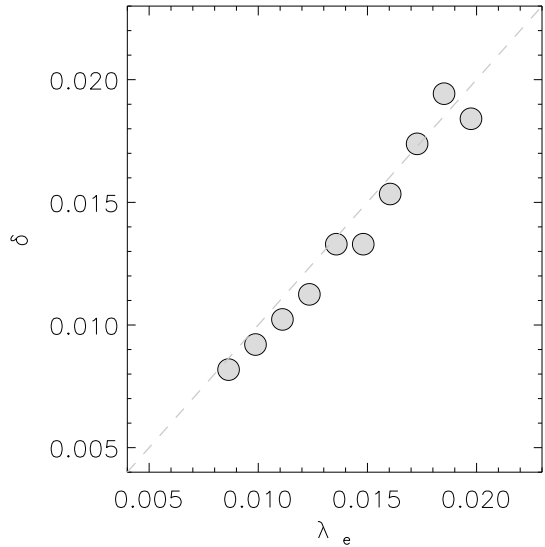}
\caption{Quasi-stationary values of $\delta$ (gray circles) as a function of $\lambda_e$. We plot the electron inertial length $\lambda_e$ in gray-dashed line for reference.}
\label{delta}
\end{figure}

From equation \eqref{sca} we see the importance of studying whether the thickness and length of the reconnection region ($\delta$ and $\Delta$, respectively) and the magnetic field at the edge of this region ($B_{in}$) change as a function of the Hall parameter. Since our scaling was performed assuming quasi-stationary conditions, we have to take this constraint into account. The width of the reconnection region $\delta$ is defined in terms of the current density profile $j(y)$ across the layer \citep{My2010}. The value of $\delta$ is obtained from a best fit of the numerical profile to a $\text{sech}^2(y/\delta)$ function, which is consistent with the initial profile give by equation \eqref{init-cond}. To determine $B_{in}$ we simply adopt  $B_{in}=B_x(x=\pi/2,y=\pi/4-\delta)$, since our neutral point is located at $x=\pi/2,y=\pi/4$. We assume that the system evolves in a quasi-stationary fashion during a time interval such that $\delta$ and $B_{in}$ show approximately no temporal variations. The length of the reconnection region $\Delta$ was obtained from the outflow velocity profile $u_{x}(x,y=\pi/4)$ applying the incompressible condition for the plasma, i.e.

\begin{equation}
u_x^{(out)}(x=\frac{\pi}{2}+\Delta,y=\frac{\pi}{4})= \frac{\Delta}{\delta}~u_y^{(in)}(x=\frac{\pi}{2},y=\frac{\pi}{4}-\delta).
\end{equation}

\begin{figure}
\centering
\includegraphics[width=.45\textwidth]{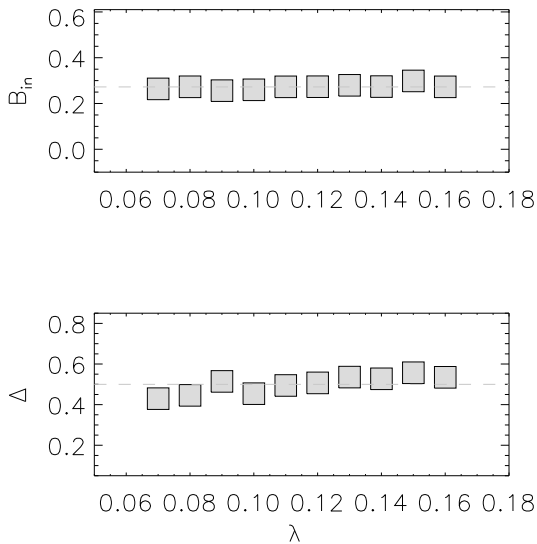}
\caption{Quasi-stationary values of $B_{in}$ (upper panel) and $\Delta$ (lower panel) as a function of $\lambda$. The gray-dashed line indicates the mean values of $B_{in}$ and $\Delta$.}
\label{DeltaBin}
\end{figure}

Figure \ref{delta} shows the quasi-stationary values of $\delta$ (gray circles) as a function of $\lambda_e$. In addition, we plot $\lambda_e$ in gray-dashed line. As expected, the width of the reconnection region is of the order of the electron inertial length. In particular, from a best linear-fit for $\log\delta-\log\lambda_e$ we obtain $\delta=(1.3\pm0.3)~\lambda_e^{1.06\pm0.07}$. Therefore, we conclude that $\delta\sim\lambda_e$.

Figure \ref{DeltaBin} shows $B_{in}$ (upper panel) and $\Delta$ (lower panel) as a function of $\lambda$ (gray squares) for the ten values of the Hall parameter. Figure \ref{DeltaBin} indicates that $B_{in}$ and $\Delta$ show approximately no dependence with the Hall parameter. This result is compatible with previous results reported in the literature \citep{S2008,W2001}. 

\begin{figure}
\centering
\includegraphics[width=.45\textwidth]{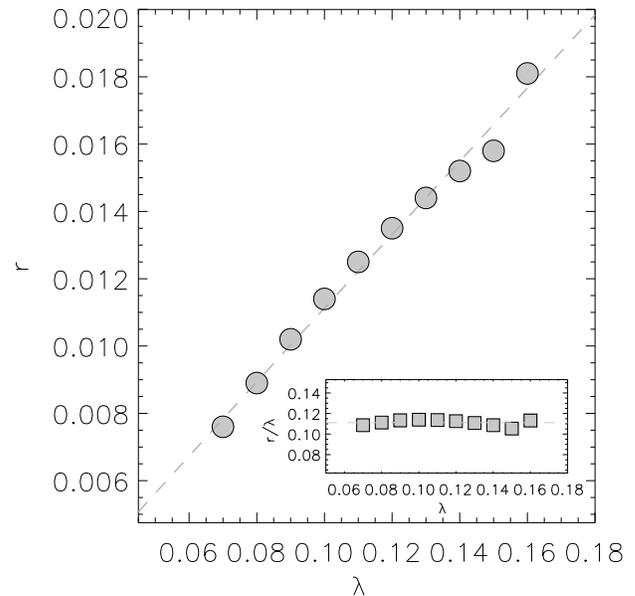}
\caption{Quasi-stationary reconnection rate $r$ (gray circles) as a function of the Hall parameter $\lambda$. The best linear-fit for $\log\lambda-\log r$ is shown in gray-dashed line. Inset: Ratio between quasi-stationary reconnection rates and the Hall parameter (gray squares) as a function of the Hall parameter.}
\label{fit}
\end{figure}

The results displayed in Figure \ref{delta} ($\delta\sim\lambda_e$) and Figure \ref{DeltaBin} ($B_{in}\sim\text{const}$ and $\Delta\sim\text{const}$) lend support to the assumptions made in equation \eqref{sca} to obtain equation \eqref{scaling}, i.e. that the reconnection rate is simply proportional to the Hall parameter. Figure \ref{fit} shows the quasi-stationary reconnection rates (gray circles), i.e. the mean reconnection rate for the time interval determined in Section \ref{stat_mag_rec}, as a function of the Hall parameter $\lambda$. In addition, we plot the curves corresponding to the best linear-fit for $\log\lambda-\log r$ (dashed line). The inset in Figure \ref{fit} shows $r/\lambda$ (gray squares) as a function of $\lambda$. From the best linear-fit for $\log\lambda-\log r$ we obtain $r=(0.11\pm0.07)~\lambda^{0.98\pm0.03}$. Therefore, we conclude that the reconnection rate $r$ is compatible with a linear relation with the Hall parameter $\lambda$, as it was predicted by our analytical relation \eqref{scaling}.

Finally, we also compare the quasi-stationary reconnection rate for a fixed value of the Hall parameter ($\lambda=0.1$), and two different electron to proton mass ratios. In particular, we compared the results for $m_e/m_i=0.015$ and $m_e/m_i=0.15$. In the quasi-stationary regime, we find approximately the same reconnection rate. This result is compatible with our theoretical result, which predicts that fast reconnection rate is insensitive to the electron to proton mass ratio even though it needs to be nonzero for reconnection to take place (Birn \textit{et al.}, 2001; Zenitani \textit{et al.}, 2011; see also Andr\'es \textit{et al.}, 2014a).

\section{Discussion and Conclusions}\label{conclus}

Within the framework of two-fluid MHD and assuming stationary conditions, we obtain a theoretical scaling for the reconnection rate. Our numerical results confirm our assumptions that the thickness of the current sheet is essentially the electron inertial length, i.e. $\delta\sim\lambda_e$, and that $B_{in}$ and $\Delta$ do not depend on the Hall parameter \citep{S2008}. More importantly, our numerical results also confirm the predicted linear dependence of the reconnection rate $r$ with the Hall parameter $\lambda$ (i.e. $r\propto\lambda$).

Within the context of incompressible HMHD, \citet{S2008} presented a quantitative analysis of reconnection valid for the resistive, HMHD and EMHD regimes. Their study concentrated on the reconnection region, without considering any particular external driving force. In the resistive MHD limit, the authors recover the standard resistive result \citep{P1957}. In the limit of EMHD, the authors find that the reconnection rate does not explicitly depend on the dissipation coefficients and features a strong dependence on the Hall parameter. In particular, they confirm an earlier result and find that $r=\sqrt{2}\lambda/\Delta$ \citep{Ch2007}, which is consistent with our scaling.

\citet{My2008} also calculated the rate of quasi-stationary, 2.5D magnetic reconnection within the framework of incompressible HMHD. The author find that the dimensionless reconnection rate is independent of the electrical resistivity and equal to $\lambda/L$, where $L$ is the scale length of the external magnetic field in the upstream region outside the electron layer. This result is also compatible with our theoretical results \citep[see also][]{My2009}.

In a different direction, \citet{W2000} reported a similar linear dependence with $\lambda$ and noted that $B_{in}$ is determined by the functional form of the boundary conditions, while $\Delta$ depends on a external time-dependent driving force. For a particular model of external driving, \citet{W2001} calculated the scaling of the reconnection rate within the framework of resistive HMHD. The authors found a $\lambda^{1/2}$ dependence for the reconnection rate. This particular scaling is not comparable with our results, since in our simulations we do not consider any external driving force. 

As discussed in the Introduction, MHD turbulence may play an important role in magnetic reconnection \citep{M1986}. \citet{Sm2004} examined the influence of the Hall effect and level of MHD turbulence on the reconnection rate in 2.5D compressible Hall MHD. Their results indicate that the reconnection rate is enhanced both by increasing the Hall parameter and by the turbulence amplitude. In agreement with these studies, our numerical results show a clear enhancement as we increase the Hall parameter. \citet{Sm2004} also suggested a power-law scaling of the reconnection rate as a function of the Hall parameter as $r\propto\delta B\lambda^{3/2}$, where $\delta B$ is the level of initial turbulence in the system. However, in our study we do not consider any initial turbulence level, since we focus on the consequences of adding the Hall effect and electron inertia terms in a laminar background. Also, in their simulations, \citet{Sm2004} added a small amount of magnetic resistivity, in order to break the frozen-in condition and start the reconnection process, which is different from our ideal EIHMHD description.

In summary, we obtained a theoretical linear scaling for the reconnection rate as a function of the Hall parameter, which is confirmed by our numerical results and is also compatible with previous results in the literature \citep{Sm2004,Ch2007,S2008,My2009,My2010}. 

%


\begin{thebibliography}{29}%
\makeatletter
\providecommand \@ifxundefined [1]{%
 \@ifx{#1\undefined}
}%
\providecommand \@ifnum [1]{%
 \ifnum #1\expandafter \@firstoftwo
 \else \expandafter \@secondoftwo
 \fi
}%
\providecommand \@ifx [1]{%
 \ifx #1\expandafter \@firstoftwo
 \else \expandafter \@secondoftwo
 \fi
}%
\providecommand \natexlab [1]{#1}%
\providecommand \enquote  [1]{``#1''}%
\providecommand \bibnamefont  [1]{#1}%
\providecommand \bibfnamefont [1]{#1}%
\providecommand \citenamefont [1]{#1}%
\providecommand \href@noop [0]{\@secondoftwo}%
\providecommand \href [0]{\begingroup \@sanitize@url \@href}%
\providecommand \@href[1]{\@@startlink{#1}\@@href}%
\providecommand \@@href[1]{\endgroup#1\@@endlink}%
\providecommand \@sanitize@url [0]{\catcode `\\12\catcode `\$12\catcode
  `\&12\catcode `\#12\catcode `\^12\catcode `\_12\catcode `\%12\relax}%
\providecommand \@@startlink[1]{}%
\providecommand \@@endlink[0]{}%
\providecommand \url  [0]{\begingroup\@sanitize@url \@url }%
\providecommand \@url [1]{\endgroup\@href {#1}{\urlprefix }}%
\providecommand \urlprefix  [0]{URL }%
\providecommand \Eprint [0]{\href }%
\providecommand \doibase [0]{http://dx.doi.org/}%
\providecommand \selectlanguage [0]{\@gobble}%
\providecommand \bibinfo  [0]{\@secondoftwo}%
\providecommand \bibfield  [0]{\@secondoftwo}%
\providecommand \translation [1]{[#1]}%
\providecommand \BibitemOpen [0]{}%
\providecommand \bibitemStop [0]{}%
\providecommand \bibitemNoStop [0]{.\EOS\space}%
\providecommand \EOS [0]{\spacefactor3000\relax}%
\providecommand \BibitemShut  [1]{\csname bibitem#1\endcsname}%
\let\auto@bib@innerbib\@empty
\bibitem [{\citenamefont {Vasyliunas}(1975)}]{V1975}%
  \BibitemOpen
  \bibfield  {author} {\bibinfo {author} {\bibfnamefont {V.~M.}\ \bibnamefont
  {Vasyliunas}},\ }\href@noop {} {\bibfield  {journal} {\bibinfo  {journal}
  {Reviews of Geophysics}\ }\textbf {\bibinfo {volume} {13}},\ \bibinfo {pages}
  {303} (\bibinfo {year} {1975})}\BibitemShut {NoStop}%
\bibitem [{\citenamefont {Dungey}(1993)}]{D1995}%
  \BibitemOpen
  \bibfield  {author} {\bibinfo {author} {\bibfnamefont {J.}~\bibnamefont
  {Dungey}},\ }\href@noop {} {\emph {\bibinfo {title} {{Physics on the
  Magnetopause}}}}\ (\bibinfo  {publisher} {AGU Monograph Vol. 90},\ \bibinfo
  {address} {AGU Washington, D. C.},\ \bibinfo {year} {1993})\ p.~\bibinfo
  {pages} {81}\BibitemShut {NoStop}%
\bibitem [{\citenamefont {Tsuneta}(1996)}]{T1996}%
  \BibitemOpen
  \bibfield  {author} {\bibinfo {author} {\bibfnamefont {S.}~\bibnamefont
  {Tsuneta}},\ }\href@noop {} {\bibfield  {journal} {\bibinfo  {journal}
  {Astrophys. J.}\ }\textbf {\bibinfo {volume} {456}},\ \bibinfo {pages} {840}
  (\bibinfo {year} {1996})}\BibitemShut {NoStop}%
\bibitem [{\citenamefont {Dungey}(2000)}]{F2007}%
  \BibitemOpen
  \bibfield  {author} {\bibinfo {author} {\bibfnamefont {J.}~\bibnamefont
  {Dungey}},\ }\href@noop {} {\emph {\bibinfo {title} {{Reconnection of
  Magnetic Fields: Magnetohydrodynamics and Collisionless Theory and
  Observations}}}}\ (\bibinfo  {publisher} {Cambridge University Press,
  Cambridge},\ \bibinfo {year} {2000})\ p.~\bibinfo {pages} {16}\BibitemShut
  {NoStop}%
\bibitem [{\citenamefont {{Parker}}(1957)}]{P1957}%
  \BibitemOpen
  \bibfield  {author} {\bibinfo {author} {\bibfnamefont {E.~N.}\ \bibnamefont
  {{Parker}}},\ }\href@noop {} {\bibfield  {journal} {\bibinfo  {journal} {J.
  Geophys. Res.}\ }\textbf {\bibinfo {volume} {62}},\ \bibinfo {pages} {509}
  (\bibinfo {year} {1957})}\BibitemShut {NoStop}%
\bibitem [{\citenamefont {{Sweet}}(1958)}]{S1958}%
  \BibitemOpen
  \bibfield  {author} {\bibinfo {author} {\bibfnamefont {P.~A.}\ \bibnamefont
  {{Sweet}}},\ }\href@noop {} {\bibfield  {journal} {\bibinfo  {journal} {The
  Observatory}\ }\textbf {\bibinfo {volume} {78}},\ \bibinfo {pages} {30}
  (\bibinfo {year} {1958})}\BibitemShut {NoStop}%
\bibitem [{\citenamefont {Yamada}(2011)}]{Y2011}%
  \BibitemOpen
  \bibfield  {author} {\bibinfo {author} {\bibfnamefont {M.}~\bibnamefont
  {Yamada}},\ }\href@noop {} {\bibfield  {journal} {\bibinfo  {journal} {Phys.
  Plasmas}\ }\textbf {\bibinfo {volume} {18}},\ \bibinfo {pages} {111212}
  (\bibinfo {year} {2011})}\BibitemShut {NoStop}%
\bibitem [{\citenamefont {{Petschek}}(1964)}]{P1964}%
  \BibitemOpen
  \bibfield  {author} {\bibinfo {author} {\bibfnamefont {H.~E.}\ \bibnamefont
  {{Petschek}}},\ }\href@noop {} {\bibfield  {journal} {\bibinfo  {journal}
  {NASA Special Publication}\ }\textbf {\bibinfo {volume} {50}},\ \bibinfo
  {pages} {425} (\bibinfo {year} {1964})}\BibitemShut {NoStop}%
\bibitem [{\citenamefont {Biskamp}(1986)}]{Bi1986}%
  \BibitemOpen
  \bibfield  {author} {\bibinfo {author} {\bibfnamefont {D.}~\bibnamefont
  {Biskamp}},\ }\href@noop {} {\bibfield  {journal} {\bibinfo  {journal}
  {Physics of Fluids}\ }\textbf {\bibinfo {volume} {29}},\ \bibinfo {pages}
  {1520} (\bibinfo {year} {1986})}\BibitemShut {NoStop}%
\bibitem [{\citenamefont {{Cai}}\ \emph {et~al.}(1994)\citenamefont {{Cai}},
  \citenamefont {{Ding}},\ and\ \citenamefont {{Lee}}}]{Ca1994}%
  \BibitemOpen
  \bibfield  {author} {\bibinfo {author} {\bibfnamefont {H.~J.}\ \bibnamefont
  {{Cai}}}, \bibinfo {author} {\bibfnamefont {D.~Q.}\ \bibnamefont {{Ding}}}, \
  and\ \bibinfo {author} {\bibfnamefont {L.~C.}\ \bibnamefont {{Lee}}},\
  }\href@noop {} {\bibfield  {journal} {\bibinfo  {journal} {J. Geophys. Res.}\
  }\textbf {\bibinfo {volume} {99}},\ \bibinfo {pages} {35} (\bibinfo {year}
  {1994})}\BibitemShut {NoStop}%
\bibitem [{\citenamefont {{Hesse}}\ \emph {et~al.}(1995)\citenamefont
  {{Hesse}}, \citenamefont {{Winske}},\ and\ \citenamefont
  {{Kuznetsova}}}]{He1995}%
  \BibitemOpen
  \bibfield  {author} {\bibinfo {author} {\bibfnamefont {M.}~\bibnamefont
  {{Hesse}}}, \bibinfo {author} {\bibfnamefont {D.}~\bibnamefont {{Winske}}}, \
  and\ \bibinfo {author} {\bibfnamefont {M.~M.}\ \bibnamefont {{Kuznetsova}}},\
  }\href@noop {} {\bibfield  {journal} {\bibinfo  {journal} {J. Geophys. Res.}\
  }\textbf {\bibinfo {volume} {100}},\ \bibinfo {pages} {21815} (\bibinfo
  {year} {1995})}\BibitemShut {NoStop}%
\bibitem [{\citenamefont {Biskamp}\ \emph {et~al.}(1997)\citenamefont
  {Biskamp}, \citenamefont {Schwarz},\ and\ \citenamefont {Drake}}]{Bi1997}%
  \BibitemOpen
  \bibfield  {author} {\bibinfo {author} {\bibfnamefont {D.}~\bibnamefont
  {Biskamp}}, \bibinfo {author} {\bibfnamefont {E.}~\bibnamefont {Schwarz}}, \
  and\ \bibinfo {author} {\bibfnamefont {J.~F.}\ \bibnamefont {Drake}},\
  }\href@noop {} {\bibfield  {journal} {\bibinfo  {journal} {Physics of
  Plasmas}\ }\textbf {\bibinfo {volume} {4}},\ \bibinfo {pages} {1002}
  (\bibinfo {year} {1997})}\BibitemShut {NoStop}%
\bibitem [{\citenamefont {{Hesse}}\ and\ \citenamefont
  {{Winske}}(1998)}]{He1998}%
  \BibitemOpen
  \bibfield  {author} {\bibinfo {author} {\bibfnamefont {M.}~\bibnamefont
  {{Hesse}}}\ and\ \bibinfo {author} {\bibfnamefont {D.}~\bibnamefont
  {{Winske}}},\ }\href@noop {} {\bibfield  {journal} {\bibinfo  {journal} {J.
  Geophys. Res.}\ }\textbf {\bibinfo {volume} {103}},\ \bibinfo {pages} {26479}
  (\bibinfo {year} {1998})}\BibitemShut {NoStop}%
\bibitem [{\citenamefont {{Shay}}\ \emph {et~al.}(1998)\citenamefont {{Shay}},
  \citenamefont {{Drake}}, \citenamefont {{Denton}},\ and\ \citenamefont
  {{Biskamp}}}]{S1998}%
  \BibitemOpen
  \bibfield  {author} {\bibinfo {author} {\bibfnamefont {M.~A.}\ \bibnamefont
  {{Shay}}}, \bibinfo {author} {\bibfnamefont {J.~F.}\ \bibnamefont {{Drake}}},
  \bibinfo {author} {\bibfnamefont {R.~E.}\ \bibnamefont {{Denton}}}, \ and\
  \bibinfo {author} {\bibfnamefont {D.}~\bibnamefont {{Biskamp}}},\ }\href@noop
  {} {\bibfield  {journal} {\bibinfo  {journal} {J. Geophys. Res.}\ }\textbf
  {\bibinfo {volume} {103}},\ \bibinfo {pages} {9165} (\bibinfo {year}
  {1998})}\BibitemShut {NoStop}%
\bibitem [{\citenamefont {Shay}\ \emph {et~al.}(2007)\citenamefont {Shay},
  \citenamefont {Drake},\ and\ \citenamefont {Swisdak}}]{Sh2007}%
  \BibitemOpen
  \bibfield  {author} {\bibinfo {author} {\bibfnamefont {M.~A.}\ \bibnamefont
  {Shay}}, \bibinfo {author} {\bibfnamefont {J.~F.}\ \bibnamefont {Drake}}, \
  and\ \bibinfo {author} {\bibfnamefont {M.}~\bibnamefont {Swisdak}},\
  }\href@noop {} {\bibfield  {journal} {\bibinfo  {journal} {Phys. Rev. Lett.}\
  }\textbf {\bibinfo {volume} {99}},\ \bibinfo {pages} {155002} (\bibinfo
  {year} {2007})}\BibitemShut {NoStop}%
\bibitem [{\citenamefont {Andr\'es}\ \emph
  {et~al.}(2014{\natexlab{a}})\citenamefont {Andr\'es}, \citenamefont {Martin},
  \citenamefont {Dmitruk},\ and\ \citenamefont {G\'omez}}]{A2014a}%
  \BibitemOpen
  \bibfield  {author} {\bibinfo {author} {\bibfnamefont {N.}~\bibnamefont
  {Andr\'es}}, \bibinfo {author} {\bibfnamefont {L.~N.}\ \bibnamefont
  {Martin}}, \bibinfo {author} {\bibfnamefont {P.}~\bibnamefont {Dmitruk}}, \
  and\ \bibinfo {author} {\bibfnamefont {D.~O.}\ \bibnamefont {G\'omez}},\
  }\href@noop {} {\bibfield  {journal} {\bibinfo  {journal} {Phys. Plasmas}\
  }\textbf {\bibinfo {volume} {21}},\ \bibinfo {pages} {072904} (\bibinfo
  {year} {2014}{\natexlab{a}})}\BibitemShut {NoStop}%
\bibitem [{\citenamefont {Birn}\ \emph {et~al.}(2001)\citenamefont {Birn},
  \citenamefont {Shay}, \citenamefont {Hesse}, \citenamefont {Kuznetsova},
  \citenamefont {Ma}, \citenamefont {Bhattacharjee}, \citenamefont {Otto},\
  and\ \citenamefont {Pritchett}}]{Bi2001}%
  \BibitemOpen
  \bibfield  {author} {\bibinfo {author} {\bibfnamefont {J.}~\bibnamefont
  {Birn}}, \bibinfo {author} {\bibfnamefont {J.~F. D. M.~A.}\ \bibnamefont
  {Shay}}, \bibinfo {author} {\bibfnamefont {N.~R. E. D.~M.}\ \bibnamefont
  {Hesse}}, \bibinfo {author} {\bibfnamefont {M.}~\bibnamefont {Kuznetsova}},
  \bibinfo {author} {\bibfnamefont {Z.~W.}\ \bibnamefont {Ma}}, \bibinfo
  {author} {\bibfnamefont {A.}~\bibnamefont {Bhattacharjee}}, \bibinfo {author}
  {\bibfnamefont {A.}~\bibnamefont {Otto}}, \ and\ \bibinfo {author}
  {\bibfnamefont {P.~L.}\ \bibnamefont {Pritchett}},\ }\href@noop {} {\bibfield
   {journal} {\bibinfo  {journal} {J. Geophys. Res.}\ }\textbf {\bibinfo
  {volume} {106}},\ \bibinfo {pages} {3715} (\bibinfo {year}
  {2001})}\BibitemShut {NoStop}%
\bibitem [{\citenamefont {{Hesse}}\ \emph {et~al.}(2001)\citenamefont
  {{Hesse}}, \citenamefont {{Birn}},\ and\ \citenamefont
  {{Kuznetsova}}}]{H2001}%
  \BibitemOpen
  \bibfield  {author} {\bibinfo {author} {\bibfnamefont {M.}~\bibnamefont
  {{Hesse}}}, \bibinfo {author} {\bibfnamefont {J.}~\bibnamefont {{Birn}}}, \
  and\ \bibinfo {author} {\bibfnamefont {M.}~\bibnamefont {{Kuznetsova}}},\
  }\href@noop {} {\bibfield  {journal} {\bibinfo  {journal} {J. Geophys. Res.}\
  }\textbf {\bibinfo {volume} {106}},\ \bibinfo {pages} {3721} (\bibinfo {year}
  {2001})}\BibitemShut {NoStop}%
\bibitem [{\citenamefont {{Pritchett}}(2001)}]{P2001}%
  \BibitemOpen
  \bibfield  {author} {\bibinfo {author} {\bibfnamefont {P.~L.}\ \bibnamefont
  {{Pritchett}}},\ }\href@noop {} {\bibfield  {journal} {\bibinfo  {journal}
  {J. Geophys. Res.}\ }\textbf {\bibinfo {volume} {106}},\ \bibinfo {pages}
  {3783} (\bibinfo {year} {2001})}\BibitemShut {NoStop}%
\bibitem [{\citenamefont {{Shay}}\ \emph {et~al.}(2001)\citenamefont {{Shay}},
  \citenamefont {{Drake}}, \citenamefont {{Rogers}},\ and\ \citenamefont
  {{Denton}}}]{S2001}%
  \BibitemOpen
  \bibfield  {author} {\bibinfo {author} {\bibfnamefont {M.~A.}\ \bibnamefont
  {{Shay}}}, \bibinfo {author} {\bibfnamefont {J.~F.}\ \bibnamefont {{Drake}}},
  \bibinfo {author} {\bibfnamefont {B.~N.}\ \bibnamefont {{Rogers}}}, \ and\
  \bibinfo {author} {\bibfnamefont {R.~E.}\ \bibnamefont {{Denton}}},\
  }\href@noop {} {\bibfield  {journal} {\bibinfo  {journal} {J. Geophys. Res.}\
  }\textbf {\bibinfo {volume} {106}},\ \bibinfo {pages} {3759} (\bibinfo {year}
  {2001})}\BibitemShut {NoStop}%
\bibitem [{\citenamefont {{Birn}}\ and\ \citenamefont
  {{Hesse}}(2001)}]{BH2001}%
  \BibitemOpen
  \bibfield  {author} {\bibinfo {author} {\bibfnamefont {J.}~\bibnamefont
  {{Birn}}}\ and\ \bibinfo {author} {\bibfnamefont {M.}~\bibnamefont
  {{Hesse}}},\ }\href@noop {} {\bibfield  {journal} {\bibinfo  {journal} {J.
  Geophys. Res.}\ }\textbf {\bibinfo {volume} {106}},\ \bibinfo {pages} {3737}
  (\bibinfo {year} {2001})}\BibitemShut {NoStop}%
\bibitem [{\citenamefont {{Otto}}(2001)}]{O2001}%
  \BibitemOpen
  \bibfield  {author} {\bibinfo {author} {\bibfnamefont {A.}~\bibnamefont
  {{Otto}}},\ }\href@noop {} {\bibfield  {journal} {\bibinfo  {journal} {J.
  Geophys. Res.}\ }\textbf {\bibinfo {volume} {106}},\ \bibinfo {pages} {3751}
  (\bibinfo {year} {2001})}\BibitemShut {NoStop}%
\bibitem [{\citenamefont {{Ma}}\ and\ \citenamefont
  {{Bhattacharjee}}(2001)}]{Ma2001}%
  \BibitemOpen
  \bibfield  {author} {\bibinfo {author} {\bibfnamefont {Z.~W.}\ \bibnamefont
  {{Ma}}}\ and\ \bibinfo {author} {\bibfnamefont {A.}~\bibnamefont
  {{Bhattacharjee}}},\ }\href@noop {} {\bibfield  {journal} {\bibinfo
  {journal} {J. Geophys. Res.}\ }\textbf {\bibinfo {volume} {106}},\ \bibinfo
  {pages} {3773} (\bibinfo {year} {2001})}\BibitemShut {NoStop}%
\bibitem [{\citenamefont {{Kuznetsova}}\ \emph {et~al.}(2001)\citenamefont
  {{Kuznetsova}}, \citenamefont {{Hesse}},\ and\ \citenamefont
  {{Winske}}}]{K2001}%
  \BibitemOpen
  \bibfield  {author} {\bibinfo {author} {\bibfnamefont {M.~M.}\ \bibnamefont
  {{Kuznetsova}}}, \bibinfo {author} {\bibfnamefont {M.}~\bibnamefont
  {{Hesse}}}, \ and\ \bibinfo {author} {\bibfnamefont {D.}~\bibnamefont
  {{Winske}}},\ }\href@noop {} {\bibfield  {journal} {\bibinfo  {journal} {J.
  Geophys. Res.}\ }\textbf {\bibinfo {volume} {106}},\ \bibinfo {pages} {3799}
  (\bibinfo {year} {2001})}\BibitemShut {NoStop}%
\bibitem [{\citenamefont {{Shay}}\ \emph {et~al.}(1999)\citenamefont {{Shay}},
  \citenamefont {{Drake}},\ and\ \citenamefont {{Rogers}}}]{S1999}%
  \BibitemOpen
  \bibfield  {author} {\bibinfo {author} {\bibfnamefont {M.~A.}\ \bibnamefont
  {{Shay}}}, \bibinfo {author} {\bibfnamefont {J.~F.}\ \bibnamefont {{Drake}}},
  \ and\ \bibinfo {author} {\bibfnamefont {B.~N.}\ \bibnamefont {{Rogers}}},\
  }\href@noop {} {\bibfield  {journal} {\bibinfo  {journal} {Geophys. Res.
  Lett.}\ }\textbf {\bibinfo {volume} {26}},\ \bibinfo {pages} {2163} (\bibinfo
  {year} {1999})}\BibitemShut {NoStop}%
\bibitem [{\citenamefont {{Morales}}\ \emph
  {et~al.}(2005{\natexlab{a}})\citenamefont {{Morales}}, \citenamefont
  {{Dasso}},\ and\ \citenamefont {{G{\'o}mez}}}]{Moa2005}%
  \BibitemOpen
  \bibfield  {author} {\bibinfo {author} {\bibfnamefont {L.~F.}\ \bibnamefont
  {{Morales}}}, \bibinfo {author} {\bibfnamefont {S.}~\bibnamefont {{Dasso}}},
  \ and\ \bibinfo {author} {\bibfnamefont {D.~O.}\ \bibnamefont
  {{G{\'o}mez}}},\ }\href@noop {} {\bibfield  {journal} {\bibinfo  {journal}
  {Journal of Geophysical Research (Space Physics)}\ }\textbf {\bibinfo
  {volume} {110}},\ \bibinfo {pages} {4204} (\bibinfo {year}
  {2005}{\natexlab{a}})}\BibitemShut {NoStop}%
\bibitem [{\citenamefont {{Morales}}\ \emph
  {et~al.}(2005{\natexlab{b}})\citenamefont {{Morales}}, \citenamefont
  {{Dasso}}, \citenamefont {{G{\'o}mez}},\ and\ \citenamefont
  {{Mininni}}}]{Mob2005}%
  \BibitemOpen
  \bibfield  {author} {\bibinfo {author} {\bibfnamefont {L.~F.}\ \bibnamefont
  {{Morales}}}, \bibinfo {author} {\bibfnamefont {S.}~\bibnamefont {{Dasso}}},
  \bibinfo {author} {\bibfnamefont {D.~O.}\ \bibnamefont {{G{\'o}mez}}}, \ and\
  \bibinfo {author} {\bibfnamefont {P.}~\bibnamefont {{Mininni}}},\ }\href@noop
  {} {\bibfield  {journal} {\bibinfo  {journal} {Journal of Atmospheric and
  Solar-Terrestrial Physics}\ }\textbf {\bibinfo {volume} {67}},\ \bibinfo
  {pages} {1821} (\bibinfo {year} {2005}{\natexlab{b}})}\BibitemShut {NoStop}%
\bibitem [{\citenamefont {Simakov}\ and\ \citenamefont
  {Chac\'on}(2008)}]{S2008}%
  \BibitemOpen
  \bibfield  {author} {\bibinfo {author} {\bibfnamefont {A.~N.}\ \bibnamefont
  {Simakov}}\ and\ \bibinfo {author} {\bibfnamefont {L.}~\bibnamefont
  {Chac\'on}},\ }\href@noop {} {\bibfield  {journal} {\bibinfo  {journal}
  {Phys. Rev. Lett.}\ }\textbf {\bibinfo {volume} {101}},\ \bibinfo {pages}
  {105003} (\bibinfo {year} {2008})}\BibitemShut {NoStop}%
\bibitem [{\citenamefont {Matthaeus}\ and\ \citenamefont
  {Lamkin}(1986)}]{M1986}%
  \BibitemOpen
  \bibfield  {author} {\bibinfo {author} {\bibfnamefont {W.~H.}\ \bibnamefont
  {Matthaeus}}\ and\ \bibinfo {author} {\bibfnamefont {S.~L.}\ \bibnamefont
  {Lamkin}},\ }\href@noop {} {\bibfield  {journal} {\bibinfo  {journal} {Phys.
  Fluids}\ }\textbf {\bibinfo {volume} {29}},\ \bibinfo {pages} {2513}
  (\bibinfo {year} {1986})}\BibitemShut {NoStop}%
\bibitem [{\citenamefont {{Lazarian}}\ and\ \citenamefont
  {{Vishniac}}(1999)}]{L1999}%
  \BibitemOpen
  \bibfield  {author} {\bibinfo {author} {\bibfnamefont {A.}~\bibnamefont
  {{Lazarian}}}\ and\ \bibinfo {author} {\bibfnamefont {E.~T.}\ \bibnamefont
  {{Vishniac}}},\ }\href@noop {} {\bibfield  {journal} {\bibinfo  {journal}
  {ApJ}\ }\textbf {\bibinfo {volume} {517}},\ \bibinfo {pages} {700} (\bibinfo
  {year} {1999})}\BibitemShut {NoStop}%
\bibitem [{\citenamefont {Smith}\ \emph {et~al.}(2004)\citenamefont {Smith},
  \citenamefont {Ghosh}, \citenamefont {Dmitruk},\ and\ \citenamefont
  {Matthaeus}}]{Sm2004}%
  \BibitemOpen
  \bibfield  {author} {\bibinfo {author} {\bibfnamefont {D.}~\bibnamefont
  {Smith}}, \bibinfo {author} {\bibfnamefont {S.}~\bibnamefont {Ghosh}},
  \bibinfo {author} {\bibfnamefont {P.}~\bibnamefont {Dmitruk}}, \ and\
  \bibinfo {author} {\bibfnamefont {W.}~\bibnamefont {Matthaeus}},\ }\href@noop
  {} {\bibfield  {journal} {\bibinfo  {journal} {Geophysical Research Letters}\
  }\textbf {\bibinfo {volume} {31}},\ \bibinfo {pages} {L02805} (\bibinfo
  {year} {2004})}\BibitemShut {NoStop}%
\bibitem [{\citenamefont {Servidio}\ \emph {et~al.}(2009)\citenamefont
  {Servidio}, \citenamefont {Matthaeus}, \citenamefont {Shay}, \citenamefont
  {Cassak},\ and\ \citenamefont {Dmitruk}}]{Se2009}%
  \BibitemOpen
  \bibfield  {author} {\bibinfo {author} {\bibfnamefont {S.}~\bibnamefont
  {Servidio}}, \bibinfo {author} {\bibfnamefont {W.~H.}\ \bibnamefont
  {Matthaeus}}, \bibinfo {author} {\bibfnamefont {M.~A.}\ \bibnamefont {Shay}},
  \bibinfo {author} {\bibfnamefont {P.~A.}\ \bibnamefont {Cassak}}, \ and\
  \bibinfo {author} {\bibfnamefont {P.}~\bibnamefont {Dmitruk}},\ }\href@noop
  {} {\bibfield  {journal} {\bibinfo  {journal} {Phys. Rev. Lett.}\ }\textbf
  {\bibinfo {volume} {102}},\ \bibinfo {pages} {115003} (\bibinfo {year}
  {2009})}\BibitemShut {NoStop}%
\bibitem [{\citenamefont {Wang}\ \emph {et~al.}(2000)\citenamefont {Wang},
  \citenamefont {Bhattacharjee},\ and\ \citenamefont {Ma}}]{W2000}%
  \BibitemOpen
  \bibfield  {author} {\bibinfo {author} {\bibfnamefont {A.}~\bibnamefont
  {Wang}}, \bibinfo {author} {\bibfnamefont {A.}~\bibnamefont {Bhattacharjee}},
  \ and\ \bibinfo {author} {\bibfnamefont {Z.~W.}\ \bibnamefont {Ma}},\
  }\href@noop {} {\bibfield  {journal} {\bibinfo  {journal} {J. Geophys. Res.}\
  }\textbf {\bibinfo {volume} {105}},\ \bibinfo {pages} {27633} (\bibinfo
  {year} {2000})}\BibitemShut {NoStop}%
\bibitem [{\citenamefont {Wang}\ \emph {et~al.}(2001)\citenamefont {Wang},
  \citenamefont {Bhattacharjee},\ and\ \citenamefont {Ma}}]{W2001}%
  \BibitemOpen
  \bibfield  {author} {\bibinfo {author} {\bibfnamefont {A.}~\bibnamefont
  {Wang}}, \bibinfo {author} {\bibfnamefont {A.}~\bibnamefont {Bhattacharjee}},
  \ and\ \bibinfo {author} {\bibfnamefont {Z.~W.}\ \bibnamefont {Ma}},\
  }\href@noop {} {\bibfield  {journal} {\bibinfo  {journal} {Phys. Rev. Lett.}\
  }\textbf {\bibinfo {volume} {87}},\ \bibinfo {pages} {265003} (\bibinfo
  {year} {2001})}\BibitemShut {NoStop}%
\bibitem [{\citenamefont {Cowley}(1985)}]{C1985}%
  \BibitemOpen
  \bibfield  {author} {\bibinfo {author} {\bibfnamefont {S.}~\bibnamefont
  {Cowley}},\ }in\ \href {\doibase 10.1007/978-94-009-5482-3_5} {\emph
  {\bibinfo {booktitle} {Solar System Magnetic Fields}}},\ \bibinfo {series}
  {Geophysics and Astrophysics Monographs}, Vol.~\bibinfo {volume} {28},\
  \bibinfo {editor} {edited by\ \bibinfo {editor} {\bibfnamefont
  {E.}~\bibnamefont {Priest}}}\ (\bibinfo  {publisher} {Springer Netherlands},\
  \bibinfo {year} {1985})\ pp.\ \bibinfo {pages} {121--155}\BibitemShut
  {NoStop}%
\bibitem [{\citenamefont {Malyshkin}(2008)}]{My2008}%
  \BibitemOpen
  \bibfield  {author} {\bibinfo {author} {\bibfnamefont {L.~M.}\ \bibnamefont
  {Malyshkin}},\ }\href@noop {} {\bibfield  {journal} {\bibinfo  {journal}
  {Phys. Rev. Lett.}\ }\textbf {\bibinfo {volume} {101}},\ \bibinfo {pages}
  {225001} (\bibinfo {year} {2008})}\BibitemShut {NoStop}%
\bibitem [{\citenamefont {Andr\'es}\ \emph
  {et~al.}(2014{\natexlab{b}})\citenamefont {Andr\'es}, \citenamefont
  {Gonzalez}, \citenamefont {Martin}, \citenamefont {Dmitruk},\ and\
  \citenamefont {G\'omez}}]{A2014b}%
  \BibitemOpen
  \bibfield  {author} {\bibinfo {author} {\bibfnamefont {N.}~\bibnamefont
  {Andr\'es}}, \bibinfo {author} {\bibfnamefont {C.}~\bibnamefont {Gonzalez}},
  \bibinfo {author} {\bibfnamefont {L.~N.}\ \bibnamefont {Martin}}, \bibinfo
  {author} {\bibfnamefont {P.}~\bibnamefont {Dmitruk}}, \ and\ \bibinfo
  {author} {\bibfnamefont {D.~O.}\ \bibnamefont {G\'omez}},\ }\href@noop {}
  {\bibfield  {journal} {\bibinfo  {journal} {Phys. Plasmas}\ }\textbf
  {\bibinfo {volume} {21}},\ \bibinfo {pages} {122305} (\bibinfo {year}
  {2014}{\natexlab{b}})}\BibitemShut {NoStop}%
\bibitem [{\citenamefont {{G{\'o}mez}}\ \emph {et~al.}(2008)\citenamefont
  {{G{\'o}mez}}, \citenamefont {{Mahajan}},\ and\ \citenamefont
  {{Dmitruk}}}]{G2008}%
  \BibitemOpen
  \bibfield  {author} {\bibinfo {author} {\bibfnamefont {D.~O.}\ \bibnamefont
  {{G{\'o}mez}}}, \bibinfo {author} {\bibfnamefont {S.~M.}\ \bibnamefont
  {{Mahajan}}}, \ and\ \bibinfo {author} {\bibfnamefont {P.}~\bibnamefont
  {{Dmitruk}}},\ }\href@noop {} {\bibfield  {journal} {\bibinfo  {journal}
  {Phys. Plasmas}\ }\textbf {\bibinfo {volume} {15}},\ \bibinfo {pages}
  {102303} (\bibinfo {year} {2008})}\BibitemShut {NoStop}%
\bibitem [{\citenamefont {{G{\'o}mez}}\ \emph {et~al.}(2013)\citenamefont
  {{G{\'o}mez}}, \citenamefont {{Martin}},\ and\ \citenamefont
  {{Dmitruk}}}]{G2013}%
  \BibitemOpen
  \bibfield  {author} {\bibinfo {author} {\bibfnamefont {D.~O.}\ \bibnamefont
  {{G{\'o}mez}}}, \bibinfo {author} {\bibfnamefont {L.~N.}\ \bibnamefont
  {{Martin}}}, \ and\ \bibinfo {author} {\bibfnamefont {P.}~\bibnamefont
  {{Dmitruk}}},\ }\href@noop {} {\bibfield  {journal} {\bibinfo  {journal}
  {Advances in Space Research}\ }\textbf {\bibinfo {volume} {51}},\ \bibinfo
  {pages} {1916} (\bibinfo {year} {2013})}\BibitemShut {NoStop}%
\bibitem [{\citenamefont {Kimura}\ and\ \citenamefont
  {Morrison}(2014)}]{K2014}%
  \BibitemOpen
  \bibfield  {author} {\bibinfo {author} {\bibfnamefont {K.}~\bibnamefont
  {Kimura}}\ and\ \bibinfo {author} {\bibfnamefont {P.~J.}\ \bibnamefont
  {Morrison}},\ }\href@noop {} {\bibfield  {journal} {\bibinfo  {journal}
  {Physics of Plasmas}\ }\textbf {\bibinfo {volume} {21}},\  (\bibinfo {year}
  {2014})}\BibitemShut {NoStop}%
\bibitem [{\citenamefont {{Morales}}\ \emph {et~al.}(2006)\citenamefont
  {{Morales}}, \citenamefont {{Dasso}}, \citenamefont {{G{\'o}mez}},\ and\
  \citenamefont {{Mininni}}}]{Mo2006}%
  \BibitemOpen
  \bibfield  {author} {\bibinfo {author} {\bibfnamefont {L.~F.}\ \bibnamefont
  {{Morales}}}, \bibinfo {author} {\bibfnamefont {S.}~\bibnamefont {{Dasso}}},
  \bibinfo {author} {\bibfnamefont {D.~O.}\ \bibnamefont {{G{\'o}mez}}}, \ and\
  \bibinfo {author} {\bibfnamefont {P.~D.}\ \bibnamefont {{Mininni}}},\
  }\href@noop {} {\bibfield  {journal} {\bibinfo  {journal} {Advances in Space
  Research}\ }\textbf {\bibinfo {volume} {37}},\ \bibinfo {pages} {1287}
  (\bibinfo {year} {2006})}\BibitemShut {NoStop}%
\bibitem [{\citenamefont {Malyshkin}(2010)}]{My2010}%
  \BibitemOpen
  \bibfield  {author} {\bibinfo {author} {\bibfnamefont {L.~M.}\ \bibnamefont
  {Malyshkin}},\ }\href@noop {} {\bibfield  {journal} {\bibinfo  {journal}
  {Phys. Scr.}\ }\textbf {\bibinfo {volume} {T142}},\ \bibinfo {pages} {8}
  (\bibinfo {year} {2010})}\BibitemShut {NoStop}%
\bibitem [{\citenamefont {Chac\'on}\ \emph {et~al.}(2007)\citenamefont
  {Chac\'on}, \citenamefont {Simakov},\ and\ \citenamefont {Zocco}}]{Ch2007}%
  \BibitemOpen
  \bibfield  {author} {\bibinfo {author} {\bibfnamefont {L.}~\bibnamefont
  {Chac\'on}}, \bibinfo {author} {\bibfnamefont {A.~N.}\ \bibnamefont
  {Simakov}}, \ and\ \bibinfo {author} {\bibfnamefont {A.}~\bibnamefont
  {Zocco}},\ }\href@noop {} {\bibfield  {journal} {\bibinfo  {journal} {Phys.
  Rev. Lett.}\ }\textbf {\bibinfo {volume} {99}},\ \bibinfo {pages} {235001}
  (\bibinfo {year} {2007})}\BibitemShut {NoStop}%
\bibitem [{\citenamefont {Malyshkin}(2009)}]{My2009}%
  \BibitemOpen
  \bibfield  {author} {\bibinfo {author} {\bibfnamefont {L.~M.}\ \bibnamefont
  {Malyshkin}},\ }\href@noop {} {\bibfield  {journal} {\bibinfo  {journal}
  {Phys. Rev. Lett.}\ }\textbf {\bibinfo {volume} {103}},\ \bibinfo {pages}
  {235004} (\bibinfo {year} {2009})}\BibitemShut {NoStop}%
\end{thebibliography}
\end{document}